\documentclass[3p,times]{elsarticle}

\usepackage{graphicx}
\usepackage{epsfig}
\usepackage{slashed}
\usepackage{dcolumn}

\begin{document}

\title{Influence of anisotropic $\Lambda/\Sigma$ creation on the $\Xi^-$ mutiplicity in
subthreshold proton--nucleus collisions}

\author[Wigner,EMMI]{Mikl\'os Z\'et\'enyi\corref{cor1}}
\ead{zetenyi.miklos@wigner.mta.hu}

\author[Wigner]{Gy\"orgy Wolf}
\ead{wolf.gyorgy@wigner.mta.hu}

\address[Wigner]{Wigner Research Centre for Physics, Hungarian Academy of Sciences, P.O. Box 49, H-1525 Budapest, 
             Hungary}
\address[EMMI]{ExtreMe Matter Institute EMMI,  
GSI Helmholtzzentrum f{\"u}r Schwerionenforschung,
  Planckstrasse 1,
  64291 Darmstadt,
  Germany}

\date{\today}

\begin{abstract}
  We present an analysis of $\Xi^-$ baryon production in subthreshold proton--nucleus ($p+A$) 
  collisions in the framework of a BUU type transport model. We propose a new mechanism
  for $\Xi$ production in the collision of a secondary $\Lambda$ or $\Sigma$ hyperon 
  and a nucleon from the target nucleus. We find that the $\Xi^-$ multiplicity in $p+A$ collisions
  is sensitive to the angular distribution of hyperon production in the primary $N+N$ collision.
  Using reasonable assumptions on the unknown elementary cross sections we are able to reproduce
  the $\Xi^-$ multiplicity and the $\Xi^-/(\Lambda+\Sigma^0)$ ratio obtained in the HADES experiment
  in $p$+Nb collisions at $\sqrt{s_{NN}} = $ 3.2~GeV energy.
\end{abstract}


\maketitle

\section{Introduction}

A sensitive tool to investigate the properties of the dense medium formed in nuclear collisions is the study of 
particle production near and below the kinematical threshold valid in an elementary $N+N$ collision. 
For the subthreshold production of a particle species, the energy has to be provided by the medium, e.g.\ via Fermi 
motion, in-medium modification of masses, or multiple collisions. The threshold for strangeness production is 
relatively high because strangeness conservation in strong interactions requires the simultaneous creation of two 
(anti)strange particles.
Therefore the study of subthreshold strangeness production can provide information about the high density medium 
reached in heavy-ion collisions at higher energies.

In hadron-nucleus ($p+A$, $\pi + A$) collisions the properties of the medium at normal nuclear matter density can 
be studied. These reactions are also an important intermediate step between hadronic and heavy-ion collisions, since 
they are cleaner, and fewer particle production channels are possible as the collision of two secondaries 
is unlikely.

The HADES collaboration has measured the yield of the doubly strange $\Xi^-$ baryon in $p$+Nb 
collisions at the subthreshold energy of $\sqrt{s_{NN}} = 3.2$~GeV \cite{Xi_pNb_HADES}, and in Ar+KCl collisions at 
the deeply subthreshold energy of $\sqrt{s_{NN}} = 2.61$~GeV \cite{Xi_ArKCl_HADES}.
They also performed a statistical model analysis of the obtained $\Xi^-$ yields along with their results for $\pi$, 
$\eta$, $\Lambda$, kaon, $\omega$ and $\phi$ multiplicities in the same reactions \cite{stat_HADES}. 
They found that the THERMUS statistical model gives a good description of all particle yields except for that 
of the $\Xi^-$ baryon, which the model underestimates by a factor 15 in the case of the reaction Ar+KCl and 
by a factor 8 in the case of the reaction $p$+Nb.

The Ar+KCl reaction was also analyzed in terms of a different statistical model in which kaon
multiplicity was taken into account on an event-by-event basis \cite{Kolomeitsev}. 
The probability to find $n$ kaons in the final state
follows a Poisson distribution. The estimate of the $\Xi$ multiplicity is based
on the observation that the doubly strange $\Xi$ baryon is produced simultaneously with two kaons, therefore
only events with at least two kaons contribute to $\Xi$ production. The value of the $\Xi/\Lambda/K^+$ multiplicity
ratio in the Ar+KCl reaction obtained in this model is smaller by a factor of 8 than the value obtained in by 
HADES and by a factor of 3 smaller than the experimental lower bound.

$\Xi$ production in heavy-ion collisions at the energies of the SIS18 synchrotron at GSI has been studied 
in terms of relativistic transport models, which are able to follow the non-equilibrium evolution of the system.
The production mechanisms of the $\Xi$ baryon in these studies include strangeness exchange reactions, in 
which the two units of strangeness of $\Xi$ are supplied by two colliding strange particles. In particular, the
reaction $\bar{K}Y \to \pi\Xi$ was considered in Ref.~\cite{rVUU1}, and the hyperon-hyperon
channel, $YY\to\Xi N$ was included in Ref.~\cite{rVUU2}. Here, and in the rest of the paper, $Y$ refers to a singly 
strange baryon, i.e.\ a $\Lambda$ or a $\Sigma$, unless otherwise stated.

All the above elementary reaction channels involve a
collision of two secondary particles, thus their role in elementary-nucleus collisions is negligible. The model
of Ref.~\cite{rVUU2} also considers the channel $\bar{K}N \to K\Xi$, which involves the exchange of two units of 
strangeness, but its contribution to the $\Xi$ yield in the Ar+KCl reaction at the HADES energy was found to be
negligible. Other ideas for $\Xi$ production channels included the reaction $\eta\Lambda \to \Xi K$ (which again
assumes the collision of two secondaries), and the excitation of a non-strange heavy resonance, which can decay
to the $\Xi K K$ final state. This latter production channel was implemented in the UrQMD transport code 
\cite{UrQMD}. In this study it was assumed that the high mass tail 
of the heavy resonances decays to the $\Xi K K$ final state with a branching ratio of 10~\%. Using this assumption
the model was able to account for the high $\Xi$ multiplicities found by HADES
in subthreshold $p$+Nb and Ar+KCl reactions.

In this paper we propose a new mechanism for the production of the $\Xi$ baryon via hyperon-nucleon collision,
$YN \to N\Xi K$. As we will see, this is particularly favorable in a hadron-nucleus reaction since it does
not involve the collision of two secondary particles. We implemented the new production mechanism in our BUU
transport code and studied $\Xi^-$ production in $p$+Nb collisions at the energy of the HADES
experiment, $\sqrt{s_{NN}} = 3.2$~GeV ($E_\textrm{kin} = 3.5$~GeV). 
$\Lambda$ and $\Sigma$ hyperons emitted in the forward direction from the first energetic collision of the projectile 
proton and a target nucleon have more energy available for $\Xi$ production in their collision with a second target 
nucleon than hyperons emitted in a different angle. Thus, the anisotropy of $\Lambda$ and $\Sigma$ production 
can influence the $\Xi$ multiplicity in $p+A$ collisions. In order to study this effect, we implemented the 
anisotropic creation of $\Lambda$ and $\Sigma$ hyperons in our BUU model.

In Section \ref{sec:YN_Xi} of this paper we discuss the new production mechanism of $\Xi$ baryons and give an 
estimate of the corresponding elementary cross section. In Section \ref{sec:aniso} we discuss the anisotropic creation
of $\Lambda$ and $\Sigma$ hyperons. We introduce our BUU model in Section \ref{sec:transport}, where we discuss
some further features of the model that have an influence on $\Xi$ production, e.g.\ in-medium hyperon potentials,
hyperon rescattering. We present and discuss our numerical results in Section \ref{sec:results}, while the conclusions
are given in Section \ref{sec:conc}. 

\section{\label{sec:YN_Xi} $\Xi$ production in hyperon-nucleon collisions}

The threshold beam kinetic energy for $\Xi$ production in proton-nucleon collisions, $p+N \to N\Xi KK$, is
$E_\textrm{kin,thr} = 3.74$~GeV, which is higher than the beam energy of $E_\textrm{kin} = 3.5$~GeV used 
in the HADES $p$+Nb experiment. The energy needed to overcome the threshold can be provided e.g.\ by 
collective effects. This means that the projectile proton interacts with more than one target nucleon (possibly 
in multiple steps) until it creates the $\Xi$ baryon. 
If e.g.\ the projectile proton interacts with two nucleons and creates a $\Xi$
via the process $p+NN \to NN\Xi KK$, the threshold beam kinetic energy reduces to 
$E_\textrm{kin,thr} = 2.56$~GeV.
This is lower than the energy of the HADES $p$+Nb experiment, showing that collective effects
can supply the energy needed for subthreshold $\Xi$ production.

In Ref.\ \cite{threebody} the role of three-body collisions in $\phi$ meson production processes 
near threshold was studied.
It was found that genuine three-body processes play a minor role in the production process as most of the cross 
section can be obtained as a sequence of two-body collisions. This factorization of three-body collisions is 
equivalent to neglecting off-shell contributions of some intermediate particles that can go on-shell in the 
given kinematics of the process. 

Encouraged by this result, we are looking for a production mechanism for the
$\Xi$ baryon via a two-step process. $\Xi$ is a doubly strange particle and it is natural to assume that the
creation of one unit of strangeness in each step of the production process is favorable over mechanisms where 
two $\bar{s}s$ quark
pairs are created in one of the two steps. These ideas lead to the production mechanism where in the first step 
a singly strange baryon, i.e.\ a $Y = \Lambda$ or $\Sigma$ hyperon is created via the reaction $p + N \to N Y K$, 
and later it collides with a second target nucleon to form a $\Xi$ via $Y + N \to N\Xi K$.

There is no experimental information on the process $Y+N \to N\Xi K$, therefore we set up a simple model 
to estimate its cross section. Our starting point is the observation that the above reaction is analogous to the 
$N+N \to NYK$ process, which has been studied in detail experimentally, and also model calculations of its 
cross section exist. Some of these calculations are based on a resonance model which assumes that hyperon 
production proceeds via an intermediate baryon resonance that decays to the $Y+K$ final state \cite{Tsushima}.

\begin{figure}[htb]
\centering
\includegraphics[width=0.45\textwidth]{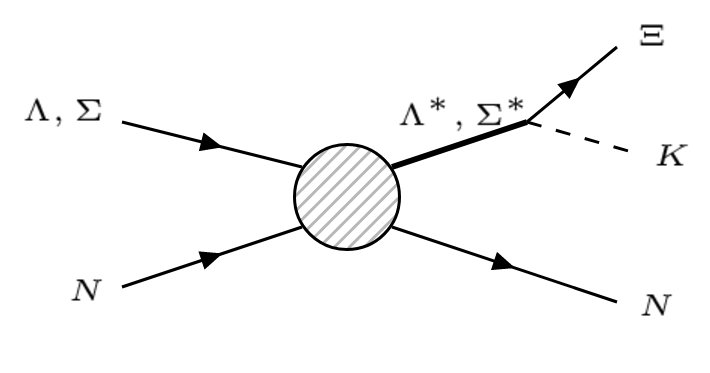}
\caption{Schematic diagram for $\Xi$ production in a hyperon-nucleon collision.}
\label{diagram}       
\end{figure}

In an analogous manner we assume that $\Xi$ production in hyperon-nucleon collisions proceeds via an intermediate 
hyperon resonance $Y^* = \Lambda^*$ or $\Sigma^*$, according to the schematic diagram of Fig.\ \ref{diagram}.
The kinematical limits for the mass of the intermediate $Y^*$ state are given by 
$m_{\Xi} + m_K < m_* < \sqrt{s}-m_N$. Here $m_*$ is the running $Y^*$ mass, which means that hyperon resonances 
with a Breit-Wigner mass slightly outside the above interval may also contribute. Based on the information given 
by the Particle Data Group (PDG) \cite{PDG}, we include 8 hyperon resonance states: $\Lambda(1820)$, $\Lambda(1830)$, 
$\Lambda(1890)$, $\Lambda(2100)$, $\Lambda(2110)$, $\Sigma(1915)$, $\Sigma(1940)$, and $\Sigma(2030)$.

The contribution of an intermediate $Y^*$ resonance to the 
cross section of the $Y+N \to N\Xi K$ process can be written in terms of the mass-differential production 
cross section of the $Y^*$ state and its mass dependent branching ratio, $BR_{Y^*\to\Xi K}(m_*) = 
\Gamma_{Y^*\to\Xi K}(m_*)/\Gamma_{\textrm{tot},Y^*}(m_*)$, as
\begin{equation}
\label{eq:sigma}
\sigma_{Y N \to N Y^* \to N\Xi K}(\sqrt{s}) = \int dm_* \frac{d\sigma_{YN\to NY^*}(\sqrt{s},m_*)}{dm_*}
\frac{\Gamma_{Y^*\to\Xi K}(m_*)}{\Gamma_{\textrm{tot},Y^*}(m_*)},
\end{equation}
where $\Gamma_{Y^*\to\Xi K}(m_*)$ and $\Gamma_{\textrm{tot},Y^*}(m_*)$ are the $\Xi K$ partial width and
the total width of a hyperon resonance $Y^*$ of actual mass $m_*$.
Assuming a constant squared matrix element, $|\mathcal{M}_{Y N\to NY^*}|^2$, for the $Y^*$ production process, 
the differential cross section of $Y^*$ production is
\begin{equation}
\label{eq:diffsigma}
\frac{d\sigma_{YN\to NY^*}(\sqrt{s},m_*)}{dm_*} = \frac{|\mathcal{M}_{Y N\to NY^*}|^2}{16\pi s}
\frac{|\mathbf{p}_f|}{|\mathbf{p}_i|} \frac{2}{N_{\textrm{pol},Y^*}}
\frac{m_*^2 \Gamma_{\textrm{tot},Y^*}(m_*)}{(m_*^2-m_{Y^*}^2)^2+m_*^2 \Gamma^2_{\textrm{tot},Y^*}(m_*)},
\end{equation}
where $m_{Y^*}$ is the Breit-Wigner mass of the hyperon resonance, $\mathbf{p}_{i(f)}$ is the incoming 
(outgoing) three-momentum in the center of momentum (CM) frame of the reaction, and $N_{\textrm{pol},Y^*}$
is the number of polarization states of the hyperon resonance $Y^*$. The mass dependence of the total 
width of the hyperon resonances is calculated as a sum of their partial widths. As the information on these 
partial widths is rather limited, we only consider the decay channels, $N\bar{K}$, $\Lambda\pi$, and 
$\Sigma\pi$ and choose values for the corresponding branching ratios based on the PDG \cite{PDG}. Although 
this choice of the branching ratios is somewhat arbitrary, the resulting uncertainty of the present rough 
estimate of the $\Xi$ production cross section is not substantial. For the mass dependent partial widths we 
use the generalized Moniz parametrization \cite{genMoniz}. Mass, width and branching ratios of hyperon resonances included in the model are listed in Table \ref{tab:Ystars}.

\begin{table}[htb]
\caption{\label{tab:Ystars} Mass, total width and branching ratios of hyperon resonances included in the model
of $\Xi$ production in hyperon-nucleon collisions.}
{\renewcommand{\arraystretch}{1.3}
\renewcommand{\tabcolsep}{0.2cm}
\begin{center}
\begin{tabular}{c|cc|ccc}
\hline
  & mass & width & \multicolumn{3}{|c}{branching ratios (\%)} \\
\cline{4-6}
 $Y^*\quad\quad J^P$ & (MeV) & (MeV) & $N\bar{K}$ & $\Lambda\pi$ & $\Sigma\pi$ \\
\hline\hline
$\Lambda(1820)\ 5/2^+$ & 1820 &  80 &  65 &  0 & 35 \\
$\Lambda(1830)\ 5/2^-$ & 1830 &  95 &  10 &  0 & 90 \\
$\Lambda(1890)\ 3/2^+$ & 1890 & 100 &  35 &  0 & 65 \\
$\Lambda(2100)\ 7/2^-$ & 2100 & 200 &  50 &  0 & 50 \\
$\Lambda(2110)\ 5/2^+$ & 2110 & 200 &  60 &  0 & 40 \\
$\Sigma(1915)\  5/2^+$ & 1915 & 120 & 100 &  0 &  0 \\
$\Sigma(1940)\  3/2^-$ & 1940 & 220 &  50 &  0 & 50 \\
$\Sigma(2030)\  7/2^+$ & 2030 & 180 &  40 & 40 & 20 \\
\hline
\end{tabular}
\end{center}
}
\end{table}

The PDG contains no information about the $\Xi K$ branching ratio of hyperon resonances. The only exception is the 
$\Lambda(2100)$ state, for which the $BR_{\Xi K} \equiv \Gamma_{Y^*\to\Xi K}/\Gamma_{\textrm{tot},Y^*}
 < 3$~\% upper limit is given \cite{PDG}. We assume that 
for each $Y^*$ state included in our model $BR_{Y^*\to\Xi K}(m_* = 2100\mathrm{\,MeV}) = 3$~\%. The constant 
squared matrix elements, $|\mathcal{M}_{Y N\to NY^*}|^2$, are treated as free parameters, and we assume that 
they have equal values for all $Y^*$ resonances. Finally, the total $\Xi$ production 
cross section is obtained as an incoherent sum of the contributions of the various $Y^*$ resonances.

We assume that the relative magnitude of the cross sections of different isospin channels follow from isospin
symmetry. This means that in order to obtain the cross section for a specific isospin channel, the isospin averaged 
cross section has to be multiplied by the appropriate isospin factor. These isospin factors depend on the 
isospin of the intermediate hyperon resonance, i.e.\ whether the diagram of Fig.\ \ref{diagram} contains a 
$\Lambda^*$ or a $\Sigma^*$ on the internal line. In this model we use the arithmetic mean of the isospin 
factors valid with an intermediate $\Lambda^*$ and $\Sigma^*$, which amounts to the assumption that the 
contribution of the two types of hyperon resonances is roughly equal.

\begin{figure}[htb]
\centering
\includegraphics[width=0.55\textwidth]{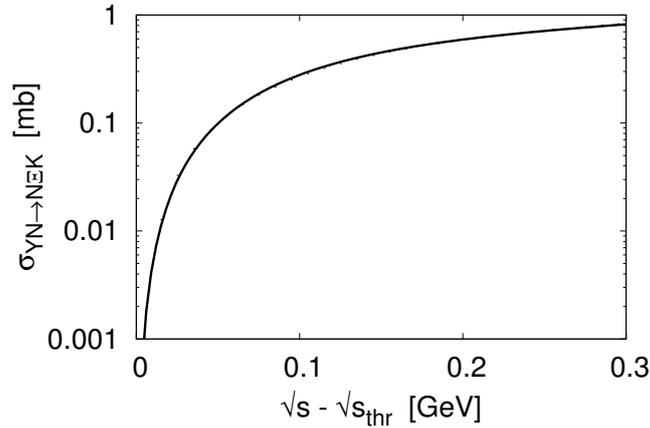}
\caption{Isospin averaged cross section of the process $Y + N \to N + \Xi + K$ ($Y = \Lambda$ or $\Sigma$) 
as a function of the excess energy.}
\label{Xi_xsec}
\end{figure}

In Fig.\ \ref{Xi_xsec}, the isospin averaged cross section of $\Xi$ production in hyperon-nucleon collisions 
is shown as a function of the excess energy, $\sqrt{s}-\sqrt{s}_\mathrm{thr}$. The curves obtained for
$\Lambda N$ and $\Sigma N$ collisions are indistinguishable. The shape of the curve is determined by our model,
while the absolute magnitude is determined by the value of the matrix element $|\mathcal{M}_{Y N\to NY^*}|^2$,
which is fixed by the requirement that the experimental $\Xi$ multiplicity in $p$+Nb reactions is reproduced, 
see Sec.\ \ref{sec:transport}. At this point we emphasize that we treat $|\mathcal{M}_{Y N\to NY^*}|^2$ as
a free parameter, therefore our model does not predict the magnitude of the $\Xi$ production cross section 
in hyperon-nucleon collisions. On the other hand we will show in Sec.\ \ref{sec:transport}
that the requirement of reproducing the experimental $\Xi$ multiplicity in $p$+Nb collisions leads to such a 
value of the matrix element $|\mathcal{M}_{Y N\to NY^*}|^2$ that gives a realistic $Y^*$ production cross
section in hyperon-nucleon scattering.

\section{\label{sec:aniso} Anisotropy of hyperon production in the $NN$ reaction}

The reaction $pp \to pK^+(\Lambda/\Sigma^0)$ has been studied by the COSY-TOF collaboration close to the 
kinematical threshold.  They also present a detailed study of the angular distribution of the three final 
state particles in different reference frames in the case of the $\Lambda$ \cite{COSY-TOF}. In particular, they 
found an increased emission of $\Lambda$ hyperons in the forward direction in the center of momentum (CM) frame 
at all three beam momenta studied in the experiment 
($p_\textrm{beam} = 2950$~MeV/$c$, 3059~MeV/$c$ and 3200~MeV/$c$).

In a $p+A$ collision, the energy available for $\Xi$ production in an elementary $Y+N$ reaction is maximal if
\textit{a)} the $Y$ hyperon is produced in the first energetic collision of the projectile proton with a 
target nucleon, and \textit{b)} the hyperon is emitted in the forward direction in the CM frame of the elementary 
$p+N$ reaction, that is,
parallel to the beam direction. This means that the increased forward emission of $\Lambda$ hyperons in $p+p$
collisions found by COSY-TOF can enhance the production of $\Xi$ baryons in subthreshold $pA$ collisions.

In Ref.~\cite{COSY-TOF}, the differential cross section of $\Lambda$ production is given in the form of an 
expansion in terms of Legendre polynomials,
\begin{equation}
  \label{eq:sigma_expansion}
  \frac{d\sigma}{d\Omega_{\Lambda}} = \sum_l a_l P_l(\cos \theta_{\Lambda}),
\end{equation}
where $\theta_{\Lambda}$ is the angle of the outgoing $\Lambda$ momentum and the beam direction in the CM frame.
While in general the $l = 0$, 1, 2, and 4 terms of the expansion are given, in the case of the $\Lambda$ baryon 
only the coefficients $a_0$ and $a_2$ are non-zero. Using $P_0 = 1$ and $P_2 = (3\cos^2\theta - 1)/2$, the 
differential cross section is given by
\begin{equation}
  \label{eq:anisotropy}
  \frac{d\sigma}{d\cos\theta_{\Lambda}} = \sigma \left[\frac{1}{2} + \frac{\xi}{4}(3\cos^2\theta_{\Lambda} - 1)
  \right],
\end{equation}
where $\sigma$ is the total cross section of the process $pp \to p\Lambda K^+$, and the anisotropy of $\Lambda$ 
production is determined by the coefficient $\xi=a_2/a_0$. Based on the positivity of the differential cross 
section we get the condition $\xi < 2$.

\begin{figure}[htb]
\centering
\includegraphics[width=0.55\textwidth]{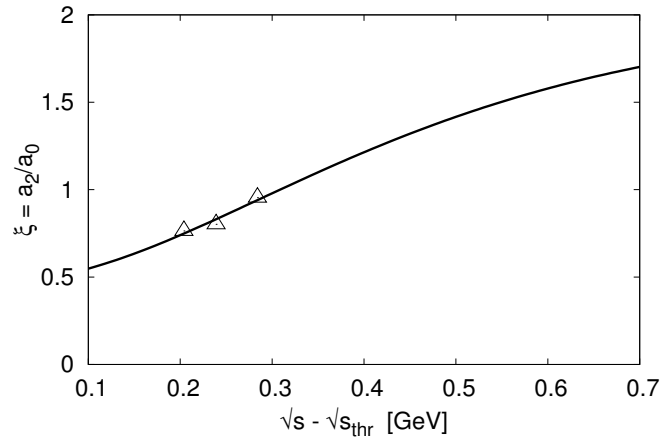}
\caption{Anisotropy coefficient $\xi=a_2/a_0$ of $\Lambda$ production in $p+p$ collisions as a function of the 
excess energy.}
\label{xi_coeff}
\end{figure}

Figure\ \ref{xi_coeff} shows the experimental values of the $\xi$ coefficient for the three energies studied by 
the COSY-TOF collaboration, as a function of the excess energy. The solid line indicates an extrapolation of 
the experimental data to higher energies.  In a $p+A$ collision 
a $\Lambda$ hyperon created in the first energetic collision of the projectile proton with a target nucleon is 
most likely to produce a $\Xi$ baryon. The kinetic energy of 3.5~GeV of the projectile proton in the HADES 
experiment corresponds to an excess energy of $\sqrt{s}-\sqrt{s}_\mathrm{thr} = 0.63$~GeV for $\Lambda$ production. 
At this energy, a value of about $\xi = 1.6$ is obtained for the anisotropy coefficient using the extrapolation 
of Fig.\ \ref{xi_coeff}.

In the extrapolation of the $\xi$ coefficient to higher energies, we have taken into account the increase of the 
experimental values 
in the 0.2~GeV$< \sqrt{s} <$0.3~GeV energy range, and assumed that $\xi$ approaches the upper limit of $\xi < 2$ with
increasing energy. 
Apart from this, the extrapolation is rather arbitrary, both higher and lower values for $\xi$ are possible at the energy 
relevant for our study. In this sense, the $\xi$ parameter can be regarded as a free parameter of the model. As we will  
see, the chosen extrapolation leads to a strong effect on the $\Xi$ multiplicity in the $p+$Nb reaction at the energy of the HADES experiment.

\section{\label{sec:transport} Transport calculations}

The early versions of the BUU code used in the present study are described in Refs.~\cite{BUU93,BUU97}.
This code was further developed in a Budapest-Rossendorf collaboration and has been successfully applied
to strangeness production near and below the production threshold \cite{BUU_strangeness,Schade} and
dilepton production \cite{BUUdilep}. 
Recently the same BUU model was used to study charmonium production in antiproton-nucleus collisions \cite{charmonium}.

Let us first remind the reader of the potentials that affect strange particle production within our BUU transport 
approach. We apply a momentum dependent mean field potential for nucleons which corresponds to a soft equation of state
with incompressibility modulus $\kappa = 215$~MeV \cite{Schade}. In line with our earlier studies of strangeness 
production \cite{BUU_strangeness,Schade}, we introduced a weakly repulsive momentum independent in-medium potential for 
kaons. This potential is linear in baryon density and yields a value of +23~MeV at normal nuclear matter density.

Furthermore, we employ in-medium potentials for $\Lambda$, $\Sigma$ and $\Xi$ baryons. In Ref.~\cite{hyperpot}
lattice QCD results for the hyperon-nucleon potential are presented, which are subsequently used to extract 
the momentum dependent single-particle potential of hyperons in nuclear medium, $U_Y(\rho,k)$. 
At zero momentum and normal nuclear matter density they obtained the values $U_{\Lambda}(\rho_0,0) = -33$~MeV, 
$U_{\Sigma}(\rho_0,0) = +11$~MeV, and $U_{\Xi}(\rho_0,0) = -6$~MeV. For $\Sigma$ and $\Xi$ baryons 
the lattice calculation found weak momentum dependence of the potential, which we neglect, 
and assume linear dependence on baryon density in such a way that the above 
values of $U_{\Sigma(\Xi)}(\rho_0,0)$ are reproduced. For the $\Lambda$ baryon
we assume a potential proportional to the nucleon in-medium potential, 
$U_{\Lambda}(\rho,k) = 0.47 \times U_N(\rho,k)$.
This form gives a good approximation of the non-negligible momentum dependence of the $\Lambda$ potential at
normal nuclear matter density, found in \cite{hyperpot}.
In-medium hyperon single particle potentials have been discussed in the context of SU(3) chiral effective
field theory in Ref.~\cite{hyperpot_EFT}. Their next-to-leading order results for the $\Lambda$ and $\Sigma$
in-medium potential are qualitatively similar to those of Ref.~\cite{hyperpot}, although their $\Lambda$
potential is less attractive at zero momentum and turns repulsive for smaller momenta, and their $\Sigma$ potential
has a stronger momentum dependence.

Next we discuss those contributions to the collision term that are relevant for $\Xi$ baryon production.
The elementary
cross sections of $\Lambda$ and $\Sigma$ production in $N+N$ collisions are implemented in the BUU code based 
on an effective Lagrangian model study, which assumes the dominance of resonance contributions \cite{Tsushima}. 
This work provides parametrizations for the energy dependence of $\Lambda$ and $\Sigma$ production cross sections
in $N+N$, $N+\Delta$ and $\Delta+\Delta$ collisions. In particular, for the reaction $pp \to p\Lambda K^+$
we get a cross section of 51~$\mu$b at the CM energy $\sqrt{s} = 3.2$~GeV
based on the parametrization of Ref.~\cite{Tsushima}. This has to be compared with the value of 38.12~$\mu$b
obtained in a partial wave analysis of the HADES $pp$ data at the same energy \cite{HADES_ppLambda}.
In order to have a realistic description of the elementary $\Lambda$ and $\Sigma$ production channels in
the BUU code, we scaled down all $NN \to NYK$ ($Y = \Lambda$ or $\Sigma$) cross sections of Ref.~\cite{Tsushima}
by a factor of 0.75. 

In Section \ref{sec:aniso} we argued that an increased forward emission of hyperons in $N+N$ reactions
can enhance the production of $\Xi$ baryons in subthreshold $pA$ collisions.
In order to investigate this effect, we implemented the anisotropy of hyperon production in $N+N$ collisions based on Eq.~(\ref{eq:anisotropy}),
where the value of the anisotropy parameter $\xi$ is taken from the extrapolation of the COSY-TOF results shown
in Fig.~\ref{xi_coeff}.
Although in Ref.~\cite{COSY-TOF} the anisotropy of only the process $pp \to N\Lambda K$ was studied, we assume the 
same angular distribution for $\Sigma$ production, $NN \to N\Sigma K$.

We considered also the absorption of $\Lambda$ and $\Sigma$ hyperons via the reaction $YN \to NN\bar{K}$. 
This process has been studied in the context of antikaon production in nuclear collisions
in Refs.~\citep{Barz_Naumann,Schade_PhD}, where also parametrizations of the relevant cross sections are given. 
We use these parametrizations in our transport calculations. 

Not only hyperon absorption, but also hyperon-nucleon elastic scattering might have an influence on $\Xi$ production,
since $\Lambda$ and $\Sigma$ hyperons loose energy even by elastic scattering and furthermore, hyperons
traveling in the forward direction might get scattered into a kinematically less 
favorable direction, thus reducing the energy available in their next collision. The inelastic reactions 
$YN \to Y^{(\prime)}N\pi$
can also have a similar effect. The cross section of the inelastic reactions can be estimated
in a way similar to the $\Xi$ production cross section described in Section \ref{sec:YN_Xi}. The energy dependence
of the total $YN \to Y^{(\prime)}N\pi$ cross section has a shape very similar to Fig.\ \ref{Xi_xsec}, reaching
values around 10~mb at $\sqrt{s_{YN}} \approx 3$~GeV, which is the maximum available CM energy for a forward
moving $Y$ hyperon.

Elastic hyperon-nucleon scattering was studied in bubble chamber experiments in the '60-s \cite{YNela1} at very low energy, 
and more recently using a scintillating fiber active target at KEK at somewhat higher energies \cite{YNela2}. Several
theoretical approaches exist for $YN$ elastic scattering which are able to reproduce the experimental data \cite{NSC,chiral}.
Both experimental results and theoretical calculations extend to hyperon momenta of $p_{\textrm{lab},Y} <~ 8$ -- 900~MeV,
while the corresponding maximal hyperon momenta in the $p$+Nb collision is around 3.6~GeV. The trends seen in the experimental 
and theoretical results suggest that the $YN$ elastic cross sections are below 10~mb at these high energies.

We implemented in the BUU the elastic $\Lambda N$ and $\Sigma N$ scattering processes with a constant cross section of 
10~mb and isotropic distribution of the final state particles in the CM frame of the reaction. This choice of the cross section
is meant to mimic also the kinematical effect of inelastic $YN \to Y^{(\prime)}N\pi$ processes.

Finally, for the $Y + N \to N\Xi K$ cross section we use our estimate shown in Fig.\ \ref{Xi_xsec}. 
We chose the value of the squared matrix element of $Y^*$ production, $|\mathcal{M}_{Y N\to NY^*}|^2$ in such a way,
that the experimental $\Xi^-$ multiplicity of $2\times 10^{-4}$ obtained by HADES in $p$+Nb collisions is reproduced 
by our transport calculation. We can also determine the total cross section for $Y^*$ creation in $YN$ collisions,
$\sigma_{Y N\to NY^*}$ via integrating Eq.\ (\ref{eq:diffsigma}) over $m_*$ and summing over all hyperon resonances $Y^*$. 
The $Y^*$ creation cross section reaches its maximal value of about 30~mb at $\sqrt{s_{YN}} \approx 3.5$~GeV. (The maximum
available CM energy in a $YN$ collision in the $p$+Nb reaction at the HADES energy is about 3~GeV.) Assuming the dominance of 
resonance contributions, $Y^*$ creation contributes a major part of the total $YN$ cross section, the other important
contribution being the creation of non-strange $N^*$ and $\Delta^*$ resonances. The total proton-proton cross section
is between 40 and 50~mb in the same energy range, therefore an $Y^*$ creation cross section of about 30~mb is a reasonable
value. However, there is still some freedom in the model for varying $\sigma_{Y N\to NY^*}$. E.g.\ if some of the 
$Y^*$ resonances have stronger couplings to the $\Xi K$ channel then the $\Xi^-$ multiplicity of HADES can be reproduced 
with smaller values of $|\mathcal{M}_{Y N\to NY^*}|^2$ and also the cross section $\sigma_{Y N\to NY^*}$ will be 
smaller.

In order to increase statistics, we use the perturbative method for strange particle production in the BUU.
This means that whenever in a two-particle collision the production threshold for a strange particle species is 
overcome, the particle is created and weighted with its production probability. At the same time, the two
colliding particles (the "parents" of the strange particle) are left untouched. Furthermore, any scattering of 
perturbative particles on "normal" particles (in the present case nucleons) influences only the pertubative 
particle, and not its colliding partner. As we will see in Section \ref{sec:results}, hyperon multiplicities are
of the order of $10^-2$ or lower. This ensure that we introduce only a small error to the reaction dynamics when
we assume that nucleons are not affected by the perturbative processes. On the other hand, collisions of a perturbative particle with its parents, or any other particles that have collided with it's parents must be avoided. This is
carefully justified in our code.

\section{\label{sec:results} Results and discussion}

The numerical results for $\Lambda$, $\Sigma^0$ and $\Xi^-$ multiplicities obtained using the BUU model described in 
the previous Section are shown in Table \ref{tab:results}. Each line in the table corresponds to a simulation
of 50000 $p$+Nb events. Line a) is obtained using the full model, i.e.\ including all the features described 
in Section \ref{sec:transport}. For the multiplicity of $\Lambda + \Sigma^0$ hyperons we got a value of 0.0197.
This result was obtained using the hyperon production cross sections reduced by a factor of 0.75, as described in the 
previous Section. The $\Lambda + \Sigma^0$ multiplicity is still above (though within the error of) the experimental
value of 0.017$\pm$0.003 obtained by HADES \cite{HADES_Lambda}. The squared matrix element of $Y^*$ production 
has been chosen to reproduce the experimental
$\Xi^-$ yield of $2.0\times 10^{-4}$ obtained by the HADES collaboration in $p$+Nb collisions. Consequently, also
the multiplicity ratio $\Xi^-/(\Lambda+\Sigma^0) = 1.015\times 10^{-2}$ lies within the range of the experimental
value of $[1.2 \pm 0.3(\textrm{stat})\pm 0.4(\textrm{syst})]\times 10^{-2}$.

We estimated the statistical uncertainty of the 
transport calculation by comparing results using the same code but with different starting point
of the random number generator. The statistical errors of particle multiplicities listed in Table \ref{tab:results}
are found to be of the order of a few percent.

We carried out calculations also with some of the features of the model switched off. These results are listed
in lines b) - f) of Table \ref{tab:results}. In line a), the anisotropy of $\Lambda$ and $\Sigma$ production according
to Section \ref{sec:aniso} is turned off, in line c), the potentials for $\Lambda$ and $\Sigma$ and $\Xi$ baryons are
neglected. Line e) contains the results obtained by turning off all $\Lambda N$ and $\Sigma N$ scattering processes, 
i.e.\ both elastic scattering and hyperon absorption via $YN \to NN\bar{K}$.
Finally, line d) corresponds to switching off both the hyperon potentials and anisotropy of hyperon production, while 
line e) corresponds to switching off $\Lambda N$ and $\Sigma N$ scattering and anisotropy of hyperon production
simultaneously.

Comparing lines c) and a) we observe
that the repulsive $\Sigma$ potential slightly reduces the $\Sigma^0$ multiplicity. The momentum dependent $\Lambda$ 
potential, although attractive at zero momentum, turns repulsive at higher momenta, hence the very small reduction
of the $\Lambda$ multiplicity. The effect of hyperon potentials on the $\Lambda$ and $\Sigma$ multiplicity is very
small, of the order of a few percent (i.e.\ the same order of magnitude as the statistical uncertainty of the 
transport calculation).  
The effect of potentials on the $\Xi^-$ multiplicity is much more pronounced. Both the repulsive $\Sigma$ and $\Lambda$
potentials (through the higher effective mass) and the attractive $\Xi$ potential enhance $\Xi$ production. This effect is
about 20~\%.

A comparison of lines e) and a) shows that
rescattering and absorption of $\Lambda$ and $\Sigma$ hyperons reduces the $\Xi^-$ multiplicity by almost 40~\%.
This is because these processes reduce the number of collisions, and the available energy in the collisions, where
the creation of a $\Xi$ is kinematically possible.

Finally, the anisotropy of hyperon production increases the forward going $\Lambda$-s and $\Sigma$-s that have higher energy available in their collision with a target nucleon. The result is a huge increase in the $\Xi$ multiplicity, by a factor of about 2.5. This can be seen by comparing lines a) with b), c) with d), or e) with f), showing that the
effect of anisotropic hyperon production is independent of hyperon potentials and hyperon rescattering.

\begin{table}[htb]
\caption{\label{tab:results} Multiplicities of $\Lambda$, $\Sigma^0$ and $\Xi^-$ baryons in central 
$p(E_\textrm{kin}=3.5 \textrm{GeV})$+Nb collisions obtained using the BUU code described in the text.
The first line lists the results of the full model, while the following lines are the results of calculations
with some features of the model switched off.}
{\renewcommand{\arraystretch}{1.3}
\renewcommand{\tabcolsep}{0.2cm}
\begin{tabular}{l|ccc|ccc|c}
\hline
  & \multicolumn{3}{c|}{ hyperon multiplicity/$10^{-2}$ } & \multicolumn{3}{c|}{$\Xi$ multiplicity/$10^{-4}$} & \\
\cline{2-7}
  & $\Lambda$ & $\Sigma^0$ & $\Lambda+\Sigma^0$ & $N\Lambda$ & $N\Sigma$ & total & $\Xi/(\Lambda+\Sigma^0)$ \\
\hline\hline
a) full model                     & 1.296 & 0.673 & 1.969 & 0.317 & 1.682 & \textbf{1.999} & 1.015$\times 10^{-2}$ \\
b) no hyperon anisotropy          & 1.307 & 0.682 & 1.989 & 0.129 & 0.655 & \textbf{0.785} & 0.395$\times 10^{-2}$ \\
c) no hyperon potentials          & 1.311 & 0.694 & 2.005 & 0.248 & 1.465 & \textbf{1.713} & 0.854$\times 10^{-2}$ \\
d) no hyperon pot.\ and aniso.    & 1.323 & 0.704 & 2.027 & 0.099 & 0.539 & \textbf{0.638} & 0.315$\times 10^{-2}$ \\
e) no hyperon absorption          & 1.302 & 0.685 & 1.987 & 0.363 & 2.787 & \textbf{3.150} & 1.585$\times 10^{-2}$ \\
f) no hyperon absorp.\ and aniso. & 1.293 & 0.669 & 1.962 & 0.135 & 1.156 & \textbf{1.292} & 0.659$\times 10^{-2}$ \\
\hline
\end{tabular}
}
\end{table}

\section{\label{sec:conc} Conclusions}

We have proposed a new mechanism for the production of the doubly strange $\Xi^-$ baryon in
subthreshold proton-nucleus collisions via a reaction of a secondary $\Lambda$ or $\Sigma$ hyperon with
a target nucleon. We implemented this mechanism in our BUU transport code and studied the reaction $p$+Nb at
$\sqrt{s_{NN}}=3.2$~GeV energy, which has been experimentally investigated by the HADES collaboration. 
According to the new mechanism, both the energy needed to create the $\Xi$ and the associated kaons,
as well as the two units of strangeness are accumulated in two steps. 

The angular distribution of the first step
process (hyperon creation) has an influence on the energy available in the second step process ($\Xi$ creation
in hyperon-nucleon collision). We have found that the anisotropy of hyperon production in $N+N$ collisions
results in a strong enhancement of the $\Xi^-$ multiplicity by about a factor of 2.5 in $p$+Nb collisions at the studied subthreshold energy. We emphasize that this enhanced particle production due to the increased forward emission of
an intermediate particle is a purely kinematical effect. Similar mechanisms might work in other cases of subthreshold
particle production, and they can play a role in $A+A$ collisions too.

We have set up a simple model for the for the experimentally unknown cross section of $\Xi$ production
in hyperon-nucleon collisions. We tuned the free parameters of the model to reproduce the $\Xi$ multiplicity
found by the HADES collaboration in subthreshold $p+$Nb collisions and we have shown that the obtained 
parameters lead to realistic cross sections for hyperon resonance production in hyperon-nucleon collisions.

The new mechanism utilizes the first two energetic elementary collisions in the
$p+A$ reaction. It is clear that in a thermalised medium formed in the same reaction, 
the energy needed for $\Xi$
creation would not be available in a hyperon-nucleon collision. This is consistent with the finding that
thermal models strongly underestimates the experimentally observed $\Xi^-$ multiplicity.

Due to the lack of information about the elementary processes playing a role in the presented mechanism, our model
involves some speculations. This includes the unknown elementary cross section of the $\Xi$ production process, which
we estimated based on some speculated properties of $\Lambda^*$ and $\Sigma^*$ resonances. Also, the anisotropy
of hyperon production in $NN$ collisions is unknown at the energy relevant for our study, we extrapolated from experimental 
results at lower energy. These uncertainties are related to the phenomenology of hyperons and hyperon resonances. New 
developments in these fields would help to decide on the importance of the presented mechanisms in subthreshold $\Xi$
production. In particular, more precise knowledge about hyperon resonances, including their decay modes, or measuring
the anisotropy of hyperon production in $NN$ collisions at the higher energies relevant for the present study would 
better constrain the parameters of our model.

\section*{Acknowledgments}

This research was supported by the Hungarian OTKA Fund No. K109462 and
by the ExtreMe Matter Institute EMMI at the GSI Helmholtzzentrum f\"ur Schwerionenforschung, Darmstadt, Germany.


\end{document}